\documentclass[twocolumn,prl,aps]{revtex4}
\usepackage{graphicx}
\usepackage{setspace}

\begin{document}

\title{Nonlocality at detection and conservation of energy\\ {\small Was Einstein looking for an ``epistemic'' interpretation, a ``superdeterministic'' one, or both?}}

\author{Antoine Suarez}
\address{Center for Quantum Philosophy\\Berninastrasse 85, 8057 Zurich/Switzerland\\suarez@leman.ch, www.quantumphil.org}

\date{May 20, 2012}

\begin{abstract}

In the Solvay conference (1927) Einstein argued against the quantum  nonlocal decision at detection on the basis of a simple single-particle experiment, but thereafter he withdrew towards the more complicated 2-particle EPR argument. It has been claimed that Einstein was seeking for an ``epistemic interpretation''. In the light of a recent experiment \cite{gszgs, as12} I argue that Einstein missed an important point: One cannot have conservation of energy without nonlocality at detection. This experiment refutes also straightforwardly ``epistemic'' and ``ontic'' alternatives to quantum theory, and shows that Einstein's ``epistemicism'' entails ``superdeterminism''.

\end{abstract}


\maketitle

\begin{spacing}{0.75}
{\small
\noindent \emph{Dedicated to Nicolas Gisin,}

\noindent \emph{who has so much contributed to the insight}

\noindent \emph{that quantum effects come from outside space-time,}

\noindent \emph{on the occasion of his 60th birthday.}}
\end{spacing}
\vspace{0.4cm}

\noindent \textbf{Introduction}.\textemdash The analysis of Einstein's attempts ``to clarify his views on quantum theory'' has led Nicholas Harrigan and Robert W. Spekkens to propose the hypothesis that Einstein ``was seeking not just any completion of quantum theory, but one wherein quantum states are solely representative of our knowledge'', or using current wording, he was seeking for an ``epistemic'' interpretation of the wave function. According to Harrigan and Spekkens their hypothesis is supported by ``the circumstance of his [Einstein's] otherwise puzzling abandonment of an even simpler argument for incompleteness from 1927'' \cite{hasp}.

In contrast with the ``epistemic'' interpretations stand the ``ontic'' ones assuming that the quantum states do exist outside our minds. In the follow entities existing within space-time are referred to as ``real''; and the ``wave function'' is said to be ``complete'', if no other elements are required to explain the outcome's distributions in experiments. In this sense, the de Broglie's interpretation (particle and ``empty wave'') is ``ontic'', assumes that the quantum states are ``real'', but does not consider the ``wave function'' complete.\cite{as12}

In the light of the recent experimental demonstration of nonlocal decision  at detection \cite{gszgs, as12} and the \emph{Proceedings of the Solvay conference 1927} \cite{bv}, I argue here that although Einstein was perfectly aware that the ``collapse of the wave function'' involves nonlocality, he missed an important point: One cannot have conservation of energy, that is ``one photon one count'' (photoelectric effect), without nonlocality at detection. This may have been the reason Einstein missed ``to clarify his views of quantum theory'' after all.

Nonetheless Harrigan and Spekkens' hypothesis about Einstein's ``epistemic'' motivation holds as well: he not only rejected quantum nonlocality and considered the ``wave function'' incomplete, but renounced to endorse the ``ontic'' interpretation of de Broglie. I argue, however, that Einstein's commitment to an ``epistemic interpretation'' entails a ``superdeterministic'' explanation, and this remains the very essence of ``quantum epistemicism'' after him as well.

I conclude that the intrinsic relationship between nonlocality and conservation of energy \cite{gszgs, as12} allow us to refute straightforwardly both, ``epistemic'' and ``ontic'' alternatives to quantum theory: This theory is both ``ontic'' and ``complete'', although the quantum states cannot be considered ``real'' and the theory can be improved.
\vspace{0.3cm}

\begin{figure}[t]
\includegraphics[width=50mm]{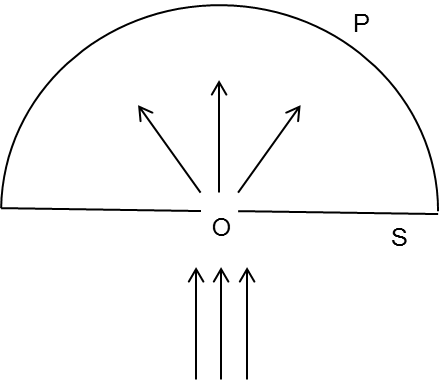}
\caption{Einstein's gedanken-experiment: Let S be a diaphragm provided with a small opening O, and P a hemispherical photographic film of large radius. Electrons impinge on S in the direction of the arrows. ``There are de Broglie waves, which impinge approximately normally on S and are diffracted at O. Behind S there are spherical waves, which reach the screen P and whose intensity at P is responsible [massgebend] for what happens at P.'' (See \cite{bv} p. 486: in the text both 'P' and 'S' are confusedly referred to as ``screen'' (\'{e}cran)).}
\label{f1}
\end{figure}

\noindent \textbf{Einstein 1927}.\textemdash In the Solvay conference Einstein considered two possible \emph{Conceptions} of the wave function in the context of the gedanken-experiment in Figure \ref{f1}:

\emph{Conception I: } The wave function does not correspond to a single electron, but to a cloud of electrons extended in space. ``According to this purely statistical point of view, $|\psi|^2$ expresses the probability that there exists at the point considered a \emph{particular} particle of the cloud, for example at a given point on the screen.''

\emph{Conception II:} The theory claims to be a complete theory of individual processes and describe everything that is governed by laws. According to this point of view, ``$|\psi|^2$ expresses the probability that at a given instant \emph{the same} particle is present at a given point (for example on the screen). (See \cite{bv} p. 486).

Einstein stresses that for reasons of coherence one has to prefer \emph{Conception II} (all information resulting from I, results also by virtue of II, but the converse is not true; and it is only by virtue of II that the conservations laws are valid for the elementary processes --among other reasons). But on the other hand Einstein has objections to make to \emph{Conception II}:
\\

\hangindent=0.36cm
\hangafter=0
\noindent
{\footnotesize ``If $|\psi|^2$ were simply regarded as the probability that at a certain point a given particle is found at a given time, it could happen that \emph{the same} elementary process produces an action in \emph{two or several} places on the screen. But the interpretation, according to which $|\psi|^2$ expresses the probability that \emph{this} particle is found at a given point, assumes an entirely peculiar mechanism of action at a distance, which prevents the wave continuously distributed in space from producing an action in two places on the screen. In my opinion, one can remove this objection only in the following way, that one does not describe the process solely by the Schr\"{o}dinger wave, but that at the same time one localises the particle during the propagation. I think that Mr de Broglie is right to search in this direction. If one works solely with the Schr\"{o}dinger waves, interpretation II of $|\psi|^2$  implies to my mind a contradiction with the postulate of relativity.''}\cite{bv}
\\

\begin{figure}[t]
\includegraphics[width=80mm]{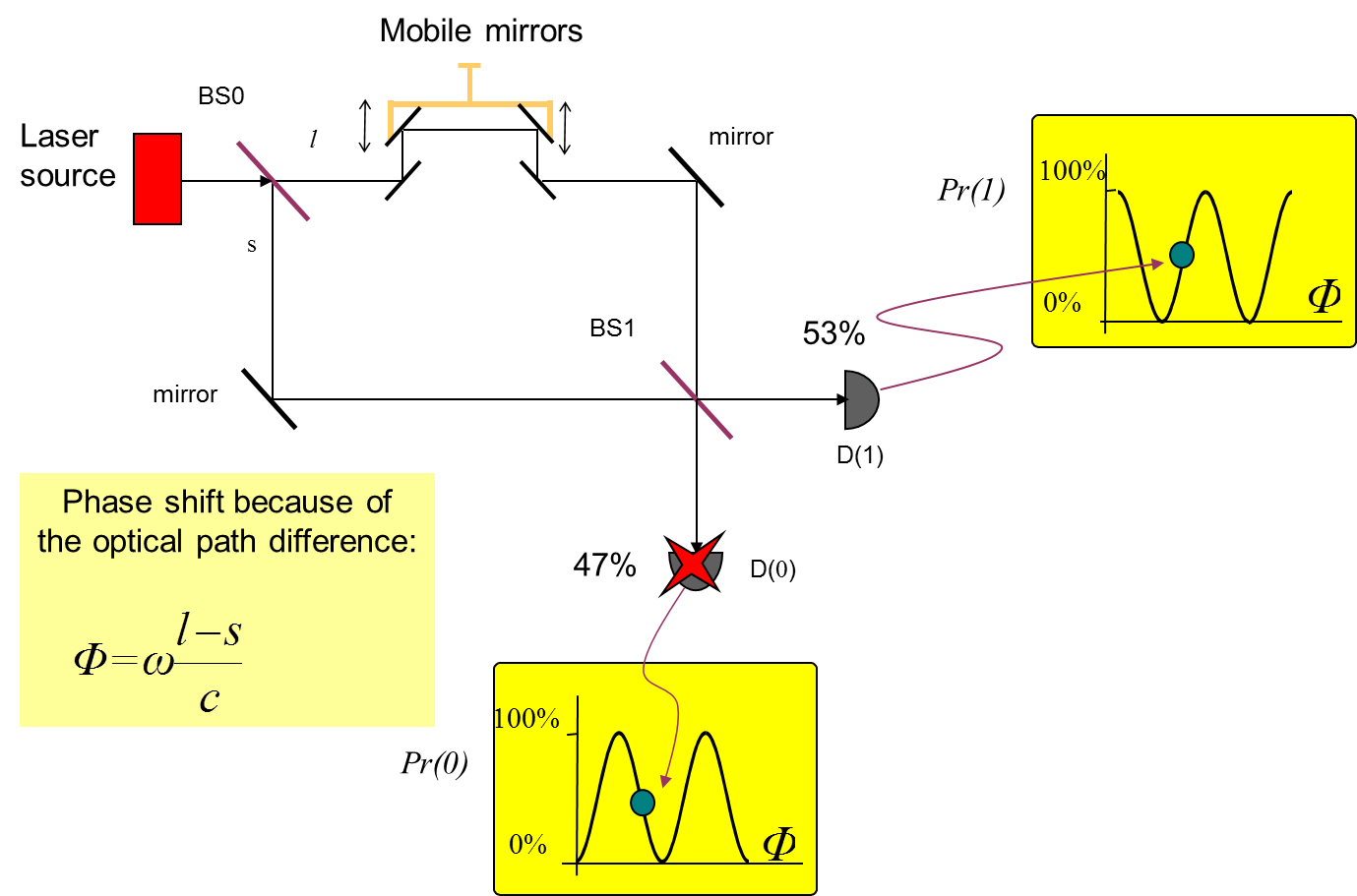}
\caption{Interference experiment: Laser light of frequency $\omega$ emitted by the source enters a Mach-Zehnder interferometer through beam-splitter (half-silvered mirror) BS0 and gets detected after leaving beam-splitter BS1. The light can reach each of the detectors D(1) and D(0) by the paths $l$ and $s$; the path-length $l$ can be changed by the experimenter. For calculating the counting rates of each detector one must take into account information about the two paths leading from the laser source to the detector (\emph{wave behavior}). However, with a single-photon source only one of the two detectors clicks: either D($1$) or D($0$) (\emph{particle behavior}): ``one photon, one count'', or conservation of energy. If $a \;\in\{+1,-1\}$ labels the detection values according to whether D($1$) or D($0$) clicks, the probability of getting $a$ is given by $P(a)=\frac{1}{2}(1+ a \cos \mathit\Phi)$, where $\mathit\Phi=\omega\tau$ is the phase parameter and $\tau=\frac{l-s}{c}$ the optical path.}
\label{f2}
\end{figure}

In the light of this statement one cannot help thinking that Einstein was a victim of ``relativistic prejudice'' and overlooked the necessity of nonlocality for the conservation of energy and the photoelectric effect. Apparently his primary intention was to oppose nonlocality. But on the basis of the gedanken-experiment he proposed (Figure \ref{f1}) nonlocality could not be questioned without questioning the conservation of energy at the same time:

It is well known that the idea of the ``collapse'' was advanced to cope with interference experiments like the sketched in Figure \ref{f2}. If the decision happens at detection and D(0) and D(1) are space-like separated, then nonlocal coordination is required in order to ensure conservation of the energy in each single quantum event (that is, avoid that sometimes both detectors fire together provoking ``one photon, two counts'', and sometimes none of them fires, provoking ``one photon, no count''). By contrast assuming decision at the beam-splitter BS1 permits to escape nonlocality in single-particle interference experiments (Figure \ref{f2}), but on the price of assuming de Broglie's ``empty waves'', that is, entities propagating within space-time that do not carry energy and momentum and are in principle inaccessible to observation.\cite{as12}

Apparently, a likely subconscious desire to fight nonlocality without giving up the conservation of energy contrived Einstein to move from decision at detection to decision at the beam-splitter. And the move brought about the more complicated EPR argument.

\vspace{0.3cm}

\noindent \textbf{The epistemic story}.\textemdash Nonetheless, it is interesting to note that in spite of EPR Einstein did not finish by endorsing ``empty waves'' (``ghost fields''). His attitude can be better understood if one puts it in relation to the two following principles:

\emph{Principle A:} All that is in space-time is accessible to observation (except in case of space-like separation).

\emph{Principle Q}: Not all that matters for physical phenomena is contained in space-time.\cite{ng}

As it appears, Einstein was reluctant regarding \emph{Principle Q} (which is the very consequence of assuming nonlocal decision at detection). But on the other hand he did not dare to reject \emph{Principle A}, and in fact he did not acknowledge ``empty waves'' as ``elements of reality''.

So Einstein's ``epistemicism'' follows from his rejection of de Broglie's ``ontic'' interpretation (with decision at the beam-splitter), and means three things: assumption of (\emph{Principle A}), adoption of decision at detection, and obviously rejection of nonlocality. But then it is sheer impossible to maintain the conservation of energy, and so the question arises: Which kind of ``reality'' underlying the ``wave function'' was Einstein looking for after all?

The only possible answer is: ``ontic states'' (underlying the quantum distributions) defined through \emph{accessible} variables in space-time. It is however easy to see, that such an explanation implies straightforwardly ``superdeterminism'':

Consider the interference experiment sketched in Figure \ref{f2}. Suppose that the quantum state entering the beam-splitter BS0 emerges from an ``ontic state'' given by certain accessible although hidden elements of reality. Then the experimenter could access information about which of the two paths ($l$ or $s$) the particles takes after leaving BS0, and thwart the interference changing the other path. This means that the view according to which the quantum distributions ``are solely representative of our knowledge'' (Einstein's ``epistemic view'') excludes any free choice on the part of the experimenter.

What about ``epistemicism'' after Einstein?

It necessarily takes the option of ``decision at detection'', otherwise it would reduce to de Broglie's ``ontic'' interpretation. Since if it were nonlocal, it would reduce to the standard quantum mechanics, it has to be local.

But if it is local, it implies violation of the ``conservation of energy''. Hence one is led to assume that the outcome is predetermined by the ``ontic state'' of the particle. This means in particular that the path the particle takes after leaving BS0 is already predetermined before the particle enters BS0, and so at the end (as we have seen before) one is led to superdeterminism.

One could object that it is not necessary to assume predetermination of the output port at BS0, but it suffices to assume that the ``ontic'' state is unobservable in principle, and the decision at BS0 (and generally at any beam-splitter) happens according to some hidden  random variable inaccessible to the experimenter (indeed this seems to be a main assumption of the PBR theorem \cite{pbr}). I give a threefold refutation to such an assumption:

- If the ``ontic'' state is unobservable in principle, then ``epistemicism'' reduces to de Broglie's ``onticism''.

- By assuming free choice on the part of the experimenter, one assumes brain outcomes ``that cannot be explained by any story in space-time''; this amount to accept \emph{Principle Q} and then there is no reason to question nonlocality and the completeness of quantum mechanics.

- Even if one assumes that at BS0 the outputport is determined by an inaccessible random variable, the experimenter could impair the quantum interference by choosing to change path $l$ and/or path $s$ at will after the particle leaves BS0. This is the argument used to rule out the assumption that the particle travels either path \emph{l} or path \emph{s} in the experiment of Figure \ref{f2}, and to conclude that for explaining interference one has to assume either nonlocal decision or de Broglie's ``empty wave'', that is get rid of ``epistemicism''. (The PBR theorem \cite{pbr} looks like a complicate version of this argument).

In conclusion: ``epistemicism'' leads necessarily to ``superdeterminism''. And as far as one wishes to avoid the latter one has to accept that there is no hidden reality underlying the ``wave function''.

Does this mean that the ``wave function'' itself is ``real'' in the sense of an entity existing within space-time? Obviously not. In the sense of \emph{Principle Q} the wave function does exist outside of our minds and it is not merely part of our knowledge, and so far is ``ontic'', but it does exist outside space-time as well, and so far is not an ``element of reality''.

Niels Bohr thinking on this issue has been repeatedly summarized through the famous quote: ``There is no quantum world. There is only an abstract physical description. It is wrong to think that the task of physics is to find out how nature is. Physics concerns what we can say about nature…'' I think that this quote is perfectly in line with the two \emph{Principles A and Q}. But independently of whether this is the right interpretation of Bohr's words, I think one should take these two principles as distinctive for quantum physics and more in general for science.
\vspace{0.3cm}

\noindent \textbf{The ontic story}.\textemdash It is interesting to go the other way around, and also analyze how things evolved from the perspective of the de Broglie's ``ontic'' interpretation.

As it is well known, David Bohm added the ``nonlocal quantum potential'' to the ``local empty wave'' to account for the nonlocal correlations in 2-particle entanglement experiments.\cite{jb} Thereafter two main developments happened leading to alternatives to quantum mechanics:

a) The ``many worlds'' interpretation attempting to restore locality. This interpretation denies both \emph{Principles A and Q}, and therefore is the consequent continuation and fulfillment of the ``empty wave'' program. It is important to be aware of the fact that if one accepts decision at the beam-splitter and ``empty waves'' one rejects (without realizing it) the \emph{Principles A and Q}, and then one will not be able to oppose ``many worlds''.

\indent b) Alternative theories that maintain nonlocality but try to weaken it through some limits. There are three testable models, which in fact have been extensively tested in the context of two particle experiments \cite{as12a}:

- \emph{Eberhard}: assumes finite-speed causal influences faster than light. Depending on the separation between Alice and Bob, there are \emph{two different states}: one defined by a set of outcome pairs exhibiting nonlocal correlations (i.e., a distribution violating Bell's inequalities, also arbitrarily large chained inequalities), and another defined by a set of outcome pairs exhibiting local ones (fulfilling Bell's inequalities) (see \cite{salart, bancal} and References therein).

- \emph{Suarez-Scarani}: assumes relativistic time-ordered nonlocal influences. Depending on whether Alice's apparatus and Bob's one move relative to each other according to the before-before relativistic timing, the model assumes two different states like Eberhard's model does (see \cite{szgs, bancal} and References therein).

- \emph{Leggett}: entails that only part of the joint outcomes of Alice and Bob (in a 2-particle experiment) are nonlocally correlated; the rest of the joint outcomes exhibit only local correlations, yielding non-trivial marginals. There is \emph{a unique state} defined by the union of the nonlocal set of outcomes and the local one, and yielding a distribution that does not violate arbitrarily large chained Bell's inequalitites (see \cite{cb, rc, bkp} and References therein).

All these three models deviate from quantum mechanics because of the limits they impose to nonlocality.

Note that \emph{Leggett} imposes a strong constraint to the union of the local and nonlocal sets of outcomes, whereas \emph{Eberhard} and \emph{Suarez-Scarani} impose none. Accordingly the last two models are neither addressed by the Colbeck-Renner's theorem nor falsified by the corresponding experiment in \cite{rc}. Therefore this experiment does not provide ``a complete answer to the question [...] of whether quantum mechanics is the optimal way to predict measurement outcomes'' (see \cite{rc} Abstract). Conversely, the theorem established in \cite{bancal} does not falsify Leggett-type models. This means that establishing completeness of quantum mechanics requires both results \cite{rc} and \cite{bancal}, provided it would be possible to achieve completion assuming decision at the beam-splitters.

However, as stated before, assuming decision at the beam-splitter amounts to accept ``many worlds'', and then one can neither incorporate the free choice of the experimenter nor nonlocality into the theory. Invoking freedom alone is not sufficient to oppose ``many worlds'', one has to reject decision at the beam-splitter as well, and accept \emph{Principle A} (space-time emerges from light and is the domain of the visible things).\cite{as12a}

Suppose now one tries to implement the ``ontic'' models under nonlocal decision at detection (which arguably \cite{as12, as12a} should be considered more basic than Bell's nonlocality \cite{jb}). De Broglie-Bohm would obviously reduce to standard quantum mechanics. As for the three models discussed previously, all of them predict uncoordinated detections at D(1) and D(0), and hence are falsified straightforwardly by the experiment in \cite{gszgs, as12}. This means that each detection outcome in a quantum experiment is nonlocal, and quantum mechanics can be considered complete, although it is susceptible of improvement (as argued in \cite{as12a}).
\vspace{0.3cm}

\noindent \textbf{Conclusion}.\textemdash The experiment presented in \cite{gszgs, as12} stresses that in the \emph{Solvay conference 1927} Einstein overlooked the relevance of quantum nonlocality for the conservation of energy. So the experiment may  contribute to overcome the still widespread prejudice that ``conservation of energy'' is nothing important for the nonlocality business. It can also help to see the \emph{Principles A and Q} as distinctive for quantum physics and science. In the light of this experiment one can finally conclude that quantum mechanics is ``ontic'' and ``complete'', although the quantum state is not ``real'' and the theory is improve-able.

\end{document}